\documentclass[a4paper,fleqn,usenatbib]{mnras}

\usepackage[T1]{fontenc}
\usepackage{ae,aecompl}

\usepackage{graphicx}	
\usepackage{amsmath}	
\usepackage{amssymb}	

\def\gtsima{$\; \buildrel > \over \sim \;$}
\def\simgt{\lower.5ex\hbox{\gtsima}}

\def\songmei2022{https://doi.org/10.48550/arXiv.2212.11034}

\title[LMC star cluster population]{A hidden link in the LMC star cluster formation history}

\author[Andr\'es E. Piatti]{
Andr\'es E. Piatti$^{1,2}$\thanks{E-mail: andres.piatti@fcen.uncu.edu.ar} \\
$^{1}$Instituto Interdisciplinario de Ciencias B\'asicas (ICB), CONICET-UNCuyo, 
Padre J. Contreras 1300, M5502JMA, Mendoza, Argentina\\
$^{2}$Consejo Nacional de Investigaciones Cient\'{\i}ficas y T\'ecnicas, Godoy Cruz 
2290, C1425FQB,  Buenos Aires, Argentina\\
}

\date{Accepted XXX. Received YYY; in original form ZZZ}

\pubyear{}

\begin{document}
\label{firstpage}
\pagerange{\pageref{firstpage}--\pageref{lastpage}}
\maketitle

\begin{abstract}
The age distribution function of star clusters in the Large Magellanic Cloud
(LMC) is known to present a feature called the cluster age gap, a period of time
from $\sim$ 4 to 11 Gyr ago with a remarkable small number of  clusters
identified. In this work we performed an in-depth analysis of three recently catalogued
age gap cluster candidates with the aim of confirming their physical nature. We
used GEMINI@GMOS $g,i$ photometry centred on the objects 
to build colour-magnitude diagrams from which we recognised their main features.
 We also performed an 
automatic search of stellar overdensities using Gaussian mixture model techniques
and analysed the corresponding colour-magnitude diagrams similarly as performed
for the aforementioned candidates. The present GEMINI data sets would seem to 
discard any strong evidence of these object being real clusters, and rather
support the possibility of the result of stellar density variations of the
LMC field star distribution. These three objects, in addition to other 17
new identified candidates, are placed in the LMC outermost disc in a limited region
toward the southwest from the LMC centre. They had embraced the possibility to
answer the long-time conundrum about the absence of LMC age gap clusters. 
From an statistical approach, combined with the knowledge of the expected LMC 
age-metallicity relationship, and recent simulations of the interaction between the
LMC and the Milky Way, we provide not only evidence against the physical nature of 
the studied objects, but also an interpretation on the lack of identification of more
LMC age gap clusters.
\end{abstract} 




\section{Introduction}

The absence of star clusters -groups of tens to thousand stars with a common 
origin in space and time- with ages between $\sim$4 and 11 Gyr in the neighbouring
galaxy, the Large Magellanic Cloud (LMC), is an astrophysical living enigma.
The LMC has more than 3740 star clusters spanning its whole lifetime; 
15 of them older than 11 Gyr \citep{betal08,piattietal2019}, while hundreds of millions of field stars 
formed during that period  \citep{hz09}. If the LMC had formed star clusters as it formed
field stars during this age gap, nearly 20 star clusters would have been found
\citep{massanaetal2022}. At present, three star clusters have been confirmed
as LMC age gap clusters, namely: ESO~121-SC03 \citep{mateoetal1986}, KMHK~1592 
\citep{piatti2022c}, and KMHK~1762 \citep{gattoetal2022}. Consequently, the absence of LMC age gap 
star clusters could imply  a particular galaxy star formation history, which strikes our 
understanding about galaxy formation and evolution processes.

\begin{figure}
\includegraphics[width=\columnwidth]{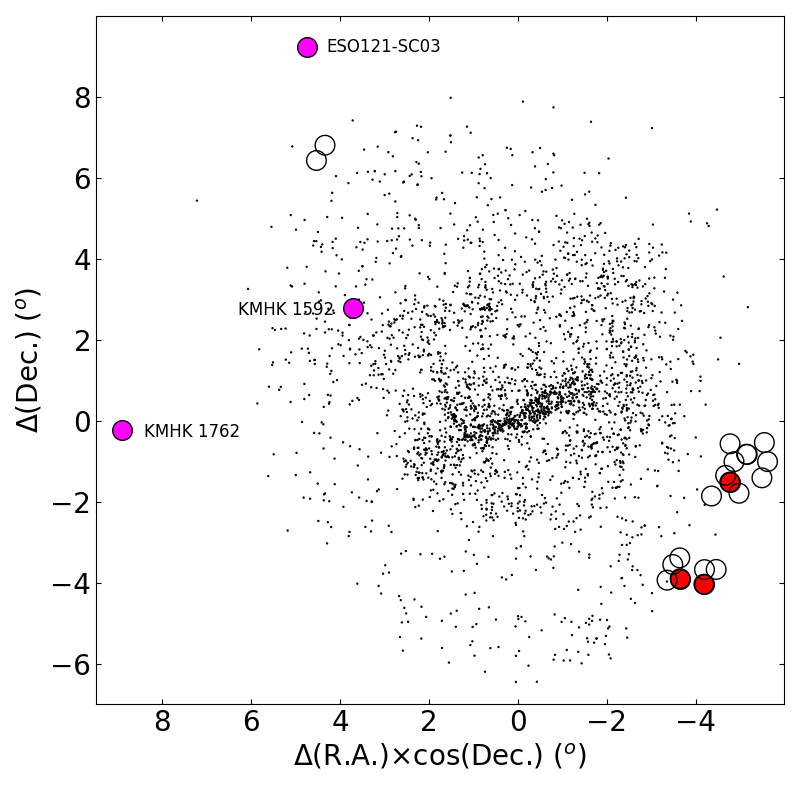}
\caption{Spatial distribution of LMC star clusters \citep[dots;][]{betal08}
and that of the objects studied in this work (red filled circles). Open and magenta
filled circles represent others \citet{gattoetal2020}'s LMC age gap candidates and
confirmed LMC age gap clusters, respectively.}
\label{fig1}
\end{figure}

Recently, \citet{gattoetal2020} discovered 20 new age gap star cluster candidates 
based on ages estimated from colour-magnitude diagrams. This result sharply contrasts
with those of different observational campaigns that have searched during the last 
decades for new LMC age gap clusters unsuccessfully \citep{dc1991,getal97}.
They also contrast with the effort of estimating ages for the vast majority of the  catalogued 
LMC clusters, using recent or ongoing 4m-class telescope/satellite imaging surveys 
throughout the LMC, e.g., SMASH \citep{nideveretal2017a}, STEP \citep{retal14}, YMCA 
\citep{gattoetal2024}, VMC \citep{cetal11}, {\it Gaia} DR3 \citep{gaiaetal2016,gaiaetal2022b}, 
etc. The resulting ages turned out to be younger than those of LMC age gap
star clusters \citep{petal14,piatti2021d,narlochetal2022}. If more
LMC age gap star clusters really exist, the above mentioned surveys
fail to detect them because of their relatively shallow photometry do not undoubtedly 
reach the age gap star cluster main sequences in the colour-magnitude diagram, 
which is mandatory to estimate star cluster ages.  
The second serious constraint is the very small number in \citet{gattoetal2020}'s candidates 
(8$\pm$2 versus more than 100 stars observed in real LMC age gap clusters;
\citet{mateoetal1986,piatti2022c}), that
could indistinguishably be the result of a composite population of LMC field
stars. Confirmed LMC age gap clusters should clearly show their main sequence 
turnoffs placed  at magnitudes compatibles with the clusters' age and distance.
The third constraint is the odd spatial distribution of \citet{gattoetal2020}'s candidates,
with clusters sharply concentrated in two areas and not distributed throughout 
the LMC.  Figure~\ref{fig1} illustrates the spatial distribution of the population of LMC star 
clusters taken from the catalogue of \citet{betal08}. The three confirmed LMC age gap star 
clusters and the new 20 ones are highlighted with large magenta and open circles, 
respectively. These constraints point to the need 
of 8m-class telescopes and high-spatial resolution imaging instruments in order 
to obtain much deeper and accurate star cluster colour-magnitude diagrams, and hence
to obtain reliable clusters' ages to confirm genuine LMC age gap clusters.

There are very few theoretical speculations 
proposed to explain the existence of the LMC cluster age gap  \citep{bekkietal2004},
and  still prevails a general lack of consensus in the community about it origin 
\citep{gattoetal2022}. The available theoretical models on the formation and
evolution of the LMC are based on a simple closed system with continuous star 
formation under the assumption of chemical homogeneity \citep{dh98},
and on a bursting formation mechanism with an important formation event 
centred at $\sim$2.0 Gyr \citep{pt1998}. The models do not predict
the interruption of cluster formation during the gap age range, so that if clusters
formed alongside the continuous field star formation, either in a closed box or bursting 
system, they should be discovered.

From recent time, the orbital motion of the LMC has been simulated with
the aim of uncovering its past interaction history 
\citep{beslaetal07,choietal2022,schmidtetal2022,jimenezarranzetal2023}.
Although these are timely efforts, the modelling of the orbital 
motions  of its star cluster population in the context of the interaction with the
Small Magellanic Cloud (SMC) and/or the Milky Way is still lacking. For instance, 
\citep{carpinteroetal2013} simulated the orbital motions of SMC star clusters with 
the aim of exploring whether the lack of SMC star clusters older than $\sim$ 7 Gyr 
could be the result of the capture of them by the LMC. Indeed, they found that star 
clusters with orbital eccentricities larger than 0.4, which represent nearly 15$\%$ 
of the SMC clusters, are captured by the LMC. Another 20-50$\%$ of the SMC clusters 
are ejected into the intergalactic medium. Similarly, some LMC clusters could have 
been stripped off the main LMC body. An example of such a possible scenario is
illustrated by a new LMC star cluster identified by \citet{p16}, which possibly
reached its present position in the outer LMC disc after being scattered from the
innermost LMC disc where it might have been born \citep{piattietal2019a}. 

The absence or the small number of age gap clusters in comparison with the
$\sim$20 expected ones will imply necessarily to consider 
improvements in the current LMC formation models.
Assuming that the LMC should harbour $\sim$ 20 age gap 
clusters, the confirmation of any fraction of \citet{gattoetal2020}'s objects as such
will definitely lead to recover a comprehensive picture of the LMC formation history 
and to improve the models of the LMC formation and evolution, including scenarios 
of tidal interaction with neighbouring galaxies through unexplored channels;
not coupled formation of clusters and field stars, etc. The larger the number of confirmed
age gap clusters, the more the reliability of current theoretical models of the LMC 
formation and evolution. The less the number of confirmed age gap 
clusters, the more the possibility of improvements in the LMC formation models, 
if they are indeed confirmed as clusters instead of LMC composite field star populations.
It is therefore necessary to definitively perform a robust statistics on the number of LMC 
age gap clusters. This is essential in order to suggest  ways of improvement to the current
LMC formation models. 

Precisely, we embarked in an observational campaign aiming at confirming the real physical
nature of these objects, and in the case of those confirmed as genuine star clusters
to estimate their fundamental parameters. The resulting ages will be used to assess
whether they are in the LMC age gap. In this work, we report results of three 
\citet{gattoetal2020}'s objects  that comprise all the observational material that we gathered
from this campaign. Although they represent, quantitatively speaking, a relative small 
percentage of the new LMC age gap star cluster candidates, the confirmation of these objects 
as age gap star clusters, or a percentage of them, or even only one of them, has the potential 
to advance our understanding of the formation  and evolution of the LMC and its dynamical 
interaction with other galaxies, among others. In Section~2 we describe the collected data 
and different procedures used to obtain the colour-magnitude diagrams. Section~3 deals with
the analysis of the colour-magnitude diagrams from the employment of different cleaning and
clustering tools, while in Section~4 we discuss on the physical nature of the studied objects
Section~5 summarizes the main conclusions of this work.

\section{Observational data}

From the sample of 20 new LMC age gap star cluster candidates (Figure~\ref{fig1}), 
STEP0004, STEP0012, and YMCA0012 were randomly selected by the GEMINI Observatory 
staff to be observed in queue mode at the GEMINI South telescope with the GMOS-S 
instrument (3$\times$1 mosaic of Hamamatsu CCDs; 5.5$\times$5.5 square 
arcmin FOV) through $g$ and $i$ filters under program GS-2024B-Q-214 (PI: Piatti).
We obtained between 5 and 8 images per filter and per object in nights with
excellent seeing (0.56$\arcsec$ to 0.97$\arcsec$ FWHM) and photometric 
conditions at a mean airmass of 1.5. We used individual exposure times of 200 sec and 79 sec 
for $g$ and $i$ filters, respectively, to prevent saturation.
We reduced the data following the recipes documented at the GEMINI Observatory 
webpage\footnote{http://www.GEMINI.edu} and employed the {\sc GEMINI@gmos} package in
{\sc GEMINI@iraf}. We applied overscan, trimming, bias subtraction, flattening and mosaicing on 
all data images, with previously obtained nightly master calibration images (bias and flats).
Then, we combined (added) all the programme images for the same object and filter with the aim 
of gaining in the reachable limiting magnitude. we found that the photometry depth increased 
nearly one mag with respect to the photometry depth reached by single programme images.

We used routines in the {\sc daophot/allstar} suite of programs \citep{setal90} to find stellar
sources and to fit their brightness profiles with  point-spread-functions (PSFs) in order to
obtain the stellar photometry of each summed image. For each of them, we obtained a preliminary 
PSF derived from the brightest, least contaminated $\sim$ 40 stars, which in turn was used to
derive a quadratically varying  PSF by fitting a larger sample of $\sim$ 100 star. Both groups 
of PSF stars were interactively selected. Then we used the {\sc allstar} programme to apply the 
resulting PSF to the identified stellar objects. {\sc allstar} generates a subtracted image which 
was used to find and measure magnitudes of additional fainter stars. We repeated this procedure 
three  times for each summed image. 
Then, we combined all the independent $g,i$ instrumental magnitudes using the stand-alone 
{\sc daomatch} and {\sc daomaster} programs\footnote{Program kindly provided by P.B. Stetson}. 
As a result, we produced one data set per object containing the $x$ and $y$ coordinates of each 
star, the instrumental $g$ and $i$ magnitudes with their respective errors, $\chi$, and
sharpness. Sharpness and $\chi$ are  image diagnostic parameters used by {\sc daophot}. With the aim 
of removing bad pixels, unresolved double stars, cosmic rays, and background galaxies from 
the photometric catalogues, we kept sources with $|$sharpness$|$ $<$ 0.5. 

\begin{figure*}
\includegraphics[width=\textwidth]{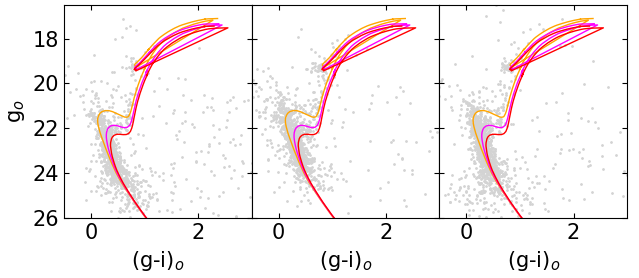}
\caption{Colour-magnitude diagrams of the observed fields centred on STEP0004 (left),
STEP0012 (middle), and YMCA0012 (right). Theoretical isochrones from 
\citet{betal12} 
of 4, 7 and 10 Gyr for [Fe/H] = -0.7 dex are superimposed with orange, magenta and 
red curves, respectively. We adopted $E(B-V)$ = 0.10 mag and $(m-M)_o$ = 18.49 mag
(see text for details).}
\label{fig2}
\end{figure*}

Figure~\ref{fig2} shows the resulting photometry for the entire GMOS FOV;
the studied  objects covering a relatively small central circular area of  radius 
0.20-0.40 arcmin \citep{gattoetal2020}. In order to build the colour-magnitude diagrams of 
Figure~\ref{fig2}, we used the instrumental $g$ magnitudes and $g-i$ colours shifted according 
to the necessary zero point offsets, which include the effect of interstellar reddening 
across the FOV ($E(B-V)$\footnote{We retrieved $E(B-V)$ values from https://irsa.ipac.caltech.edu/}
 = 0.055$\pm$0.005 mag), to match the zero-age main sequence and the red giant clump
of theoretical isochrones \citep[][PARSEC version 
2.1S\footnote{https://stev.oapd.inaf.it/cgi-bin/cmd}]{betal12} corresponding to 4, 7, and 10 Gyr,
respectively, which span the age range of the LMC age gap. Note that these offsets
do not affect the assessment of the physical reality of the studied objects, nor the
estimation of ages of genuine star clusters either.
We also adopted a mean LMC distance modulus of $(m-M)_o$ =
18.49 mag \citep{dgetal14}, and an overall metallicity of [Fe/H] = -0.7 dex, which is the 
mean metallicity value of the LMC age-metallicity relationship during the age gap period \citep{pg13}.
As can be seen, the colour-magnitude diagrams resulted to be more than 2 mag deeper
than those obtained by \citet{gattoetal2020}. They would seem to reveal the presence of 
field star populations composed of intermediate-age ones (1-2 Gyr) to moderately old stars 
($\sim$10 Gyr).

\section{Data analysis}
\subsection{Intrinsic colour-magnitude diagrams}

The presence of a star cluster in a star field implies the existence of a
local stellar overdensity with distinctive magnitude and colour distributions,
whose stars delineate the cluster sequences in the colour-magnitude diagram.
Hence, the colour-magnitude diagram observed along the cluster's line-of-sight 
contains information of both cluster and field stars. This means that if the
colour-magnitude diagram features of the star field were subtracted, the
intrinsic characteristics of the cluster colour-magnitude diagram would be uncovered.
One possibility of carrying out such a distinction consists in comparing a star field
colour-magnitude diagram with that observed along the cluster's line of sight and
then to properly eliminate from the later a number of stars equal to that found 
in the former, bearing in mind that the magnitudes and colours of the eliminated stars 
in the cluster's colour-magnitude diagram must reproduce the respective magnitude and 
colour distributions in the star field colour-magnitude diagram. The resulting 
subtracted field star colour-magnitude diagram should depict the intrinsic features
of that cleaned particular field that, in the case of being a star cluster,  are the
well known cluster's sequences.

We used a circular region around the centrers of the studied objects with a radius
equals to twice their radii, which were estimated from previously constructed stellar radial 
profiles based on star counts. We adopted for the three objects a radius of 0.40 
arcmin, in very good agreement with the radii quoted in Table~B1 of \citet{gattoetal2020}.
Similarly, we devised 1000 star field circles with the same radius of the objects' circles
and centred at four times the objects' radii. They were randomly distributed around the
objects' circles. We then built the respective 1000 colour-magnitude diagrams with the aim of
considering the variation in the stellar density and in the magnitude and colour distributions 
across the objects' neighbouring regions. The procedure to select stars to subtract from the 
object colour-magnitude diagram follows the precepts outlined by \citet{pb12}. 
We applied it using one star field colour-magnitude diagram at a time to be compared with the
object colour-magnitude diagram. It consists in defining boxes centred on the magnitude and 
colour of each star of the star field colour-magnitude diagram; then to superimpose them on 
the object colour-magnitude diagram, and finally to choose one star per box to subtract. In 
order to guarantee  that at least one star is found within the box boundary, we considered 
boxes with size of ($\Delta$$g$,$\Delta$$(g-i)$) = (1.00 mag, 0.40 mag). In the case that 
more than one star is located inside a box, the closest one to its centre is subtracted. 
During the choice of the subtracted stars we took into account their magnitude and colour 
errors by allowing them to have a thousand different values of magnitude and colour within 
an interval of $\pm$1$\sigma$, where $\sigma$ represents the errors in their magnitude and 
colour, respectively. We also imposed the condition that the spatial positions of the subtracted 
stars were chosen randomly. In practice, for each field star we randomly selected a
position in the object's circle and searched for a star to subtract within a box of
0.1$\arcmin$ a side. We iterated this loop up to 1000 times, if no star was found in the 
selected spatial box. 

\begin{figure*}
\includegraphics[width=\textwidth]{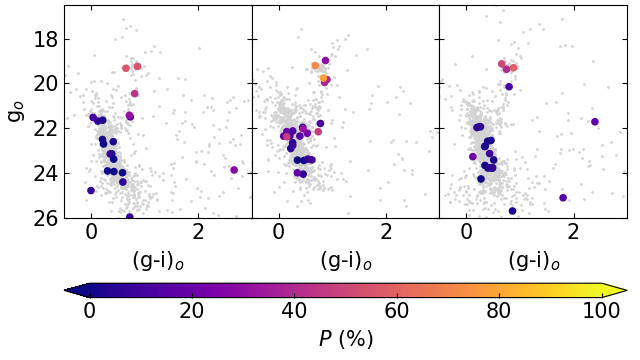}
\caption{Cleaned colour-magnitude diagrams built from the present GEMINI data of  STEP0004 (left),
STEP0012 (middle), and YMCA0012 (right). The colour bar represents (in $\%$)
how much the magnitude and colour of a star differ from the magnitude and
colour distribution of the projected surrounding LMC star field population.
The grey points represent all the measured stars.}
\label{fig3}
\end{figure*}

The outcome of the cleaning procedure is an object colour-magnitude 
diagram that likely contains only stars that represent the intrinsic features of that
region. From the 1000 different cleaned object colour-magnitude diagrams, we defined
a probability $P$ ($\%$) of being intrinsic magnitudes and colours as the ratio $N$/10, where 
$N$ (between 0 and 1000) is the number of times a star was found among the 1000 different 
cleaned colour-magnitude diagrams. Figure~\ref{fig3} shows the resulting cleaned colour-magnitude 
diagrams for the three studied objects. They include with large filled circles all the stars
measured within the objects' radii and coloured according to their respective $P$ value.

\subsection{Detection of stellar overdensities}

\begin{figure*}
\includegraphics[width=\textwidth]{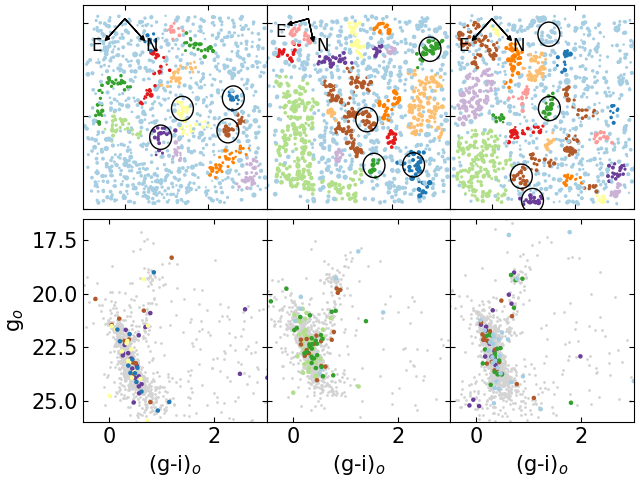}
\caption{Spatial distribution of the measured stars  from the present GEMINI data 
in the fields of STEP0004, STEP0012, 
and YMCA0012, from top left to top right panels, respectively, with the identified HDBSCAN 
stellar overdensities distinguished with different coloured symbols. The bottom panels
show the colour-magnitude diagrams of the group of stars encircled with black circles
in  the top panels, respectively.}
\label{fig4}
\end{figure*}

Instead of cleaning the object colour-magnitude diagram, which implies
assuming the existence of an intrinsic stellar overdensity, we performed
alternatively a different approach consisting in detecting spatial
clustering in the studied stellar fields. We used the HDBSCAN 
\citep[Hierarchical Density-Based Spatial Clustering of Applications with Noise,][]{campelloetal2013}
Gaussian mixture model technique \citep{hr2021} to search for spatial stellar 
overdensities in the ($x,y$) coordinates plane. The \texttt{min$\_$cluster$\_$size} 
parameter was varied between 2 and 15 dex in steps of 1 dex, and from the output we 
built the spatial stellar distribution plots  as illustrated in Fig.~\ref{fig4} (top panels).
We coloured points according to the \texttt{clusterer.labels$\_$} parameter, which labels the
different groups of stars identified by HDBSCAN.

The number of groups and the stars included in them can vary with the 
\texttt{min$\_$cluster$\_$size} parameter. Therefore, we visually inspected Figure~\ref{fig4} 
looking for the optimum \texttt{min$\_$cluster$\_$size} value towards which the
\texttt{clusterer.labels$\_$} values remain constant and with similar star distributions.  
Each \texttt{clusterer.labels$\_$} value corresponds to a particular group of stars in Figure~\ref{fig4}. HDBSCAN provides also the membership probability of each star to the corresponding group, all of them with membership probabilities higher than 90$\%$.
For comparison purposes, we built colour-magnitude diagrams for the group of stars identified 
at the centre of each observed field, which is in excellent agreement with the positions of 
the objects identified by \citet{gattoetal2020}, and three additional detected overdensities
with similar dimensions like those of the studied objects. The chosen overdensities are encircled
in the top panels of Figure~\ref{fig4}, while their colour-magnitude diagrams
are shown in the bottom panels, respectively, drawn with the corresponding colours. 

\section{Discussion}

We started examining the schematic finding charts built by \citet{gattoetal2020}
for all their studied star clusters. They distinguished stars with cluster membership
probability smaller than 25$\%$, equals to 50$\%$, and higher than 75$\%$, respectively.
As can be seen (their Figure~B1), the analysed area would seem to be nearly nine
times larger than the clusters' ones. Curiously, star with cluster membership
probabilities higher than 75$\%$ are distributed throughout the entire analysed region, 
which implies that stars with magnitudes and colours similar to those of clusters' members
are also spread outside the clusters' areas. This behaviour could mean either that: 
1) the field star cleaning colour-magnitude diagram procedure used was not able to
robustly clean the analysed regions around the clusters' areas, which weakens the
reliability of the assigned membership probabilities; or 2) stars with magnitudes and 
colours similar to those located in the clusters' areas already exist 
across the entire analysed regions, which would imply that the \citet{gattoetal2020}'s 
candidates could be the result of variation (overdensity) of the composite LMC star field 
population. Note that this effect is seen in all the studied clusters regardless 
their ages.

The LMC disc has an inclination of 34$^o$ and a position angle of the line of node of 
139.1$^o$ \citep[][]{vdmk14}. This means that the heliocentric distances of
star clusters located in the LMC outer disc are different from that of the LMC
geometrical centre. Particularly, the studied objects located at projected distances of 
$\sim$ 5$\degr$ and position angles of $\sim$ 230$\degr$ (see Figure~\ref{fig1}) could be on 
average nearly 5 kpc behind the LMC centre (eq. (1) of \citet{santosetal2020}), 
which implies a difference in their distance moduli with respect to the mean value for 
the LMC of $\sim$0.20-0.25 mag. If we used the theoretical isochrones of \citet{betal12}
for the $g$ versus $g-i$ colour-magnitude diagram, we would find that a 0.20-0.25 mag 
shift in distance modulus translates into a difference of age $\gtrapprox$ 2 Gyr,
for ages in the gap age range. \citet{gattoetal2020} adopted the mean LMC distance modulus
when fitting theoretical isochrones for all the clusters, so that their derived 
ages for the presently studied objects could be subestimated. They derived ages of 8.9,
7.1 and 8.9 Gyr for STEP0004, STEP0012, and YMCA0012, respectively, so that by adding
2 Gyr these objects would fall nearly at the upper end or outside the LMC age gap. 

\citet{gattoetal2020} derived overall metallicities for the studied objects of [M/H] 
= -0.40 dex. Combining this value with the estimated ages, the objects would seem to
fall well outside the LMC age-metallicity relation, whose average value is $\sim$ 
-0.75 dex for the gap age range \citep[][and references therein]{piatti2021g}. A mean
metallicity of [M/H] = -0.40 dex is typical of LMC stellar populations younger than
2 Gyr, which are mostly confined to the inner LMC disc and bar \citep{pg13}. Although
neither the unexpected derived metallicities, nor the adopted mean distance modulus, 
nor the assigned membership probabilities are conclusive evidence against the physical
nature of the studied objects as LMC age gap clusters, all of them considered together 
represent a volume of outcomes that discourage confidently concluding
on the studied objects as real star clusters. They would rather seem to be fluctuations
of the stellar density along the line-of-sight.

The procedure to assign probabilities $P$ described in Section~3.1 has been
tested and validated elsewhere \citep[see, e.g.,][and references therein]{pl2022,piatti2022}.
Particularly,  we applied it to KMHK~1592, an LMC star cluster projected toward a 
relatively crowded star field (see Figure~\ref{fig1}). For that purpose, we used a set of data 
obtained with the GEMINI South telescope in combination with the GMOS-S instrument and 
the $g$ and $i$ filters, similarly as the present data sets, and clearly disentangled
its sub-giant and red giant branches \citep[see][]{piatti2022c}. As a matter of providing
additional examples of the robustness of the cleaning procedure, we included in the
Appendix the analysis of a well known LMC star cluster, as well as  the corresponding
HDBSCAN results. 

Figure~\ref{fig3} shows that the measured stars placed within the adopted radii
of the studied objects are mostly main sequence stars with a probability higher than 
90$\%$ ($P$ $<$ 10$\%$) of having magnitudes and colours representative of the projected 
LMC star field populations, and a couple of red clump stars that can be either part of 
the surrounding field or a local enhancement. There is no cluster sequences composed
of stars with $P$ $>$ 70$\%$. The apparent stellar overdensities  would seem to be
the result of variations in the stellar density alone the line-of-sight. Note that
Figure~\ref{fig3} depicts all the stars enclosed within the adopted radii, 
so that the number of stars that represent the excess of stars over the mean field 
stellar density is still smaller. Because of the relatively large heliocentric
distance of the LMC, its composite field star population, or a small portion of it,
can appear distributed in the colour-magnitude diagram similarly to clusters' stars.

In order to probe that small LMC field star overdensities can mimic star cluster
colour-magnitude diagrams, we performed a blind search of stellar overdensities using
the HDBSCAN engine. By entering HDBSCAN simply with the $x$ and $y$ coordinates 
of all the measured stars in a given observed field, HDBSCAN identified tens of 
stellar overdensities of different size, shape and number of stars. Fortunately, 
HDBSCAN also identified the stellar overdensities called by \citet{gattoetal2020} 
STEP0004, STEP0012, and YMCA0012, located at the centre of their respective observed 
fields. Top panels of Figure~\ref{fig4} illustrate these findings. Using the stars 
belonging to the identified stellar overdensities, we built the colour-magnitude 
diagrams of STEP0004, STEP0012, and YMCA0012 and those of three randomly selected 
stellar overdensities per observed field with sizes similar to the studied objects. 
As can be seen in the bottom panels of Figure~\ref{fig4}, the colour-magnitude 
diagrams resemble those of star clusters with the unavoidable presence of some 
outliers. The randomly chosen stellar overdensities have not been catalogued as star 
cluster candidates. Nevertheless, we applied to their colour-magnitude diagrams the 
cleaning procedure of Section~3.1, and found that all their stars have $P$ 
$<$ 10$\%$. 

\citet{piatti2021d} applied the procedure to remove the contamination caused
by field stars in the colour-magnitude diagrams to all the LMC age gap cluster
candidates identified by \citet{gattoetal2020}, using their same data sets and
that of SMASH DR2 \citep{nideveretal2021}, to further confirm the ages of
the new star cluster candidates, so to reinforce their discoveries. He
concluded that the objects would be LMC star field density fluctuations rather 
than age gap star clusters, despite neither \citet{gattoetal2020} nor SMASH
surveys have the ability to disentangle the existence of LMC age gap star 
clusters. He pointed to the need of further deeper observations of these star 
cluster candidates with the aim of providing a definite assessment on their 
physical realities. Indeed, the present GEMINI data sets would seem to discard 
any strong evidence of STEP0004, STEP0012, and YMCA0012 being real star clusters.
Since \citet{piatti2021d} and present results are in agreement, we concluded that
the remaining LMC age gap cluster candidates with no deeper observations could
also be a group of stars with similar properties like those of the studied objects.
We note that there are other characteristics that a group of stars must comply with 
in order to be confirmed as a genuine open cluster, such as a 
stellar number density profile, a cluster mass function,  a mean metallicity, a relation 
between cluster mass and cluster radius, among others \citep{piattietal2023}.

Additionally, based only on the similar spatial distribution of the three observed 
fields and that of the remaining \citet{gattoetal2020}'s candidates (see figure~\ref{fig1}),
we found that the present results for the studied objects could be extended to
nearly 50$\%$ of the whole \citet{gattoetal2020}'s candidate sample. In order to 
obtain such a completeness
probability, we used eq. (1) of \citet{p17a}, which combines the ratio of the
number of observed fields to the total number of LMC age gap cluster candidates
and a measure of the similarity between their spatial distributions, respectively. 
Thus, the better an LMC area is covered by the analysed fields, the higher the 
similarity of their respective spatial distributions. To obtain a measure of
the spatial similarity we employed the {\it kde.test} statistical function provided 
by the {\it ks} package (version 1.10.4)\footnote{http //www.mvstat.net/tduong.}. 
We assumed the null hypothesis is true, of observing a result at least as extreme 
as the value of the test statistic \citep{fb2012}.
Here the null hypothesis is that  ($x,y$) coordinate samples for studied objects
and for the whole object sample come from the same underlying 
distribution. We obtained a similarity value of 0.95. Therefore, it would not seem
to be obvious that a substantial number of LMC age gap cluster
candidates are real star clusters.

Indeed, by looking at Figure~1 of  \citet{gattoetal2020}, the candidates
come from a limited number of tiles analysed, which in turn represent a minor
percentage of the whole survey area coverage. Their concentration 
in a relatively small region could infer the occurrence of a bursting LMC age gap
formation episode, which contrast with the spatial distribution of the three known
LMC age gap clusters, scattered across the outer LMC disc on the other
side of the galaxy (see Figure~\ref{fig1}). 

With accurate heliocentric 
distances, proper motions and radial velocities, it could be possible to trace the 
star cluster orbits backward in time for a time interval equals to their ages, in 
order to recover their birthplaces. \citep{pl2022} performed such an analysis to
conclude that the recently discovered stellar system YMCA-1 is a moderately old
SMC star cluster that could be stripped by the LMC during any of the close 
interactions between  both galaxies, and now is seen to the East from the  LMC centre.
By starting with the 
modelling of the orbital motions of ESO~121-SC03, KMHK~1592 and KMHK~1762 
--the three known LMC age gap  clusters-- we could infer whether they have been
subject of tidal stripping, so that we presently observe them in the outer LMC disc.
Indeed, \citet{vasiliev2024} found that the first close passage of the LMC by the
Milky Way could have happened between $\sim$ 5-10 Gyr ago, just along the
LMC age gap. This is a promising scenario which deserves further attention by the
next generation of numerical models of LMC cluster orbits, in order
to explain the lack of more LMC clusters with ages between $\sim$ 4 and 11 Gyr.
Because of the apparent fruitless efforts in the search for LMC age gap clusters 
from the present results, it would seen suggestive to focus further analysis
on the improvement of the LMC cluster formation history that involves galaxy
interactions.

\section{Conclusions}

The LMC cluster age gap is a phenomenon that still remains not understood in
the context of our knowledge of the LMC star cluster formation and evolution. It is related
to the lack of star clusters with ages between $\sim$ 4 and 11 Gyr as numerous
as those expected. Recently,
\citet{gattoetal2020} discovered a number of LMC age gap cluster candidates 
that matches those expectations. Because of the need of deeper 
colour-magnitude diagrams to reliably estimate their ages, we carried out 
GEMINI@GMOS observations of three of these candidates, namely:
STEP0004, STEP0012 and YMCA0012. The outcomes of 
our data analysis can be summarized as follows:\\

$\bullet$ The inspection of their colour-magnitude diagrams, once they
were cleaned of the presence of field stars, would suggest that the stars
which represent an excess over the mean field star spatial density distribution 
have in general probabilities of being cluster members smaller than 10$\%$.\\

$\bullet$ Across the observed fields we identified many other stellar 
overdensities distributed around the star cluster candidates. They have not
been catalogued as star cluster candidates. Moreover, when
we inspected their color-magnitude diagrams decontaminated from field
stars, we found in general stars with probabilities of being part of physical 
star groups smaller than 10$\%$, as expected.\\

$\bullet$ Even in the case of assuming the existence of real star clusters,
the procedure to estimate their astrophysical parameters employed by
\citet{gattoetal2020} could mislead the results. For instance, they
adopted the mean LMC heliocentric distance as the clusters' distance, which
could imply to estimate ages younger than $\sim$ 2 Gyr. The derived
metal content for these star cluster candidates put them at the same
time among the population of LMC clusters younger than $\sim$ 2 Gyr\\

$\bullet$ Because of the similar spatial distribution in the LMC of the three
studied objects and that of the remaining age gap cluster candidates of
\citet{gattoetal2020}, the present results on the presence of field star
density variations instead of the existence of real star clusters, could be 
extended to nearly 50$\%$ of the whole sample.\\

$\bullet$ The lack of detection of more LMC age gap clusters point to the 
need of exploring different cluster formation and evolution models that 
include the gravitational effects from the interaction between the involved
galaxies. Particularly, we speculate with the possibility that a number
of LMC age gap clusters could have been ejected because a
close encounter between the LMC and the Milky Way, that could
have happened according to recent theoretical simulations at that
space of time.

\section*{Acknowledgements}
We thank the referee for the thorough reading of the manuscript and
timely suggestions to improve it. 

Based on observations obtained at the international GEMINI Observatory, a program of NSF NOIRLab, which is managed by the Association of Universities for Research in Astronomy (AURA) under a cooperative agreement with the U.S. National Science Foundation on behalf of the GEMINI Observatory partnership: the U.S. National Science Foundation (United States), National Research Council (Canada), Agencia Nacional de Investigaci\'{o}n y Desarrollo (Chile), Ministerio de Ciencia, Tecnolog\'{i}a e Innovaci\'{o}n (Argentina), Minist\'{e}rio da Ci\^{e}ncia, Tecnologia, Inova\c{c}\~{o}es e Comunica\c{c}\~{o}es (Brazil), and Korea Astronomy and Space Science Institute (Republic of Korea).

\section{Data availability}

Data used in this work are available upon request to the author.






\appendix

\section{CMD analysis and stellar overdensities}

\begin{figure}
\includegraphics[width=\columnwidth]{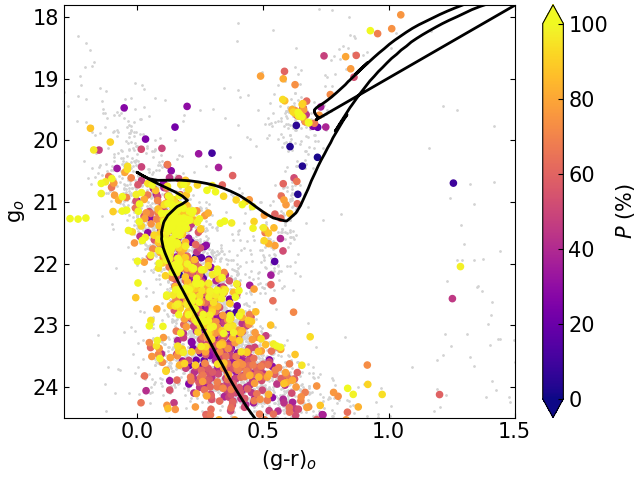}
\caption{Cleaned colour-magnitude diagrams of stars measured in
the field of SL~529 built from GEMINI data of  \citet{p13}. The colour bar 
represents (in $\%$)
how much the magnitude and colour of a star differ from the magnitude and
colour distribution of the projected surrounding LMC star field population.
The grey points represent all the measured stars.
A theoretical isochrones from  \citet{betal12} of 2.25 Gyr and [Fe/H] = -0.60 dex
is superimposed. We adopted $E(B-V)$ = 0.04 mag and $(m-M)_o$ = 18.49 mag.}
\label{fig1a}
\end{figure}

\begin{figure*}
\includegraphics[width=\textwidth]{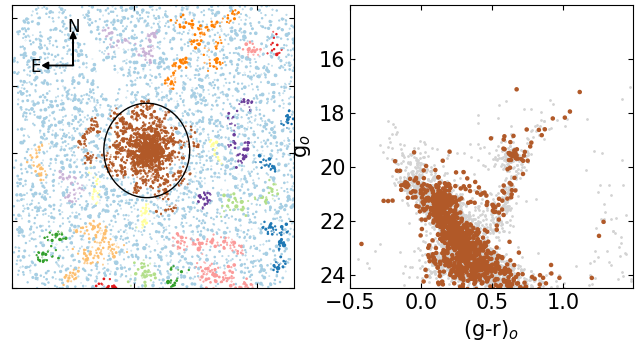}
\caption{ Spatial distribution of the measured stars in the field of SL~529
 with the identified HDBSCAN 
stellar overdensities distinguished with different coloured symbols. The right panel
shows the colour-magnitude diagram of the group of stars encircled with black circle
in the left panel.}
\label{fig2a}
\end{figure*}

With the aim of providing additional examples of the performance of the
procedures applied in Sections~3.1 and 3.2, we used GEMINI@GMOS
data obtained by \citet{p13} for the LMC star cluster SL~529. 
In Figure~\ref{fig1a} we show the CMD of the stars measured in the
cluster's field, similar to Figure~\ref{fig3}. We  superimposed
the theoretical isochrone from \citet{betal12} for the cluster's age (2.25 Gyr) 
and metallicity ([Fe/H] = -0.60 dex),
and adopting the cluster's reddening ($E(B-V)$ = 0.04 mag ) and distance
modulus (18.49 mag), for comparison purposes. In Figure~\ref{fig2a}
we show for SL~529 a figure similar to Figure~\ref{fig4}, built from the
outcomes of the HDBSCAN method.

\bsp	
\label{lastpage}
\end{document}